\documentclass[twocolumn]{aastex631} 

\usepackage{amsmath}
\usepackage{ulem}

\begin{document}

\title{Three-Dimensional Polarimetric Structure of Jets from Radio Galaxy MRC 0600-399 in A3376}

\author[0000-0002-4037-1346]{Haruka Sakemi}
\affiliation{Graduate School of Sciences and 
Technology for Innovation, Yamaguchi University, 1677-1 Yoshida, Yamaguchi, 753-0841, Japan}

\author[0000-0002-9875-7436]{James O. Chibueze}
\affiliation{Department of Mathematical Sciences, University of South Africa, Cnr Christian de Wet Rd and Pioneer Avenue, Florida Park, 1709 Roodepoort, South Africa}
\affiliation{Department of Physics and Astronomy, Faculty of Physical Sciences, University of Nigeria, Carver Building, 1 University Road, Nsukka 410001, Nigeria}

\author[0000-0001-7363-6489]{William D. Cotton}
\affiliation{National Radio Astronomy Observatory, 520 Edgemont Road, Charlottesville, VA 22903, USA}
\affiliation{South African Radio Astronomy Observatory, Liesbeek House, River Park, Gloucester Road, Cape Town, 7700, South Africa}

\author[0000-0001-6282-6025]{Viral Parekh}
\affiliation{National Radio Astronomy Observatory (NRAO), 1003 Lopezville Rd, Socorro, NM 87801, USA}

\author[0000-0002-0040-8968]{Takumi Ohmura}
\affiliation{Institute for Cosmic Ray Research, The University of Tokyo, 5-1-5 Kashiwanoha, Kashiwa, Chiba 277-8582, Japan}
\affiliation{National Astronomical Observatory of Japan, 2-21-1 Osawa, Mitaka, Tokyo 181-8588, Japan}

\author[0000-0001-6353-7639]{Mami Machida}
\affiliation{Division of Science, National Astronomical Observatory of Japan, 2-21-1 Osawa, Mitaka, Tokyo 181-8588, Japan}

\author[0000-0003-4369-7314]{Taichi Igarashi}
\affiliation{Department of Physics, Rikkyo University, Tokyo 171-8501, Japan}
\affiliation{Division of Science, National Astronomical Observatory of Japan, 2-21-1 Osawa, Mitaka, Tokyo 181-8588, Japan}

\author[0000-0001-9399-5331]{Takuya Akahori}
\affiliation{Mizusawa VLBI Observatory, National Astronomical Observatory of Japan, 2-21-1 Osawa, Mitaka, Tokyo 181-8588, Japan}

\author[0000-0003-1949-7005]{Hiroki Akamatsu}
\affiliation{SRON Netherlands Institute for Space Research, Niels Bohrweg 4, 2333 CA Leiden, The Netherlands}
\affiliation{International Centre for Quantum-field Measurement Systems for Studies of the Universe and Particles (QUP), The High Energy Accelerator Research Organization
(KEK), 1-1 Oho, Tsukuba, Ibaraki 305-0801, Japan}

\author[0009-0004-9046-6521]{Hiroyuki Nakanishi}
\affiliation{Liberal Arts Education Center, Ishikawa Prefectural University, 1-308 Suematsu, Nonoichi, Ishikawa 921-8836}

\author[0000-0001-8416-7673]{Tsutomu T.\ Takeuchi}
\affiliation{Division of Particle and Astrophysical Science, Nagoya University, Furo-cho, Chikusa-ku, Nagoya 464--8602, Japan}
\affiliation{The Research Center for Statistical Machine Learning, the Institute of Statistical Mathematics, 10-3 Midori-cho, Tachikawa, Tokyo 190---8562, Japan}



\begin{abstract}

MRC 0600-399, in the merging galaxy cluster Abell 3376, is one of the unique Head-Tail radio galaxies having abruptly bending structures, the origin of which is still under discussion. A previous study suggested that the interaction between the jets and the magnetic fields on the cold front has produced such an outstanding feature; however, it is still not known how the jet propagates the long distance keeping the sharp shape. In this study, we performed a polarization analysis of MRC 0600-399 and a nearby radio source using MeerKAT L-band data. To interpret the Faraday structure, we constructed the Pseudo-3D visualization map of the reconstructed Faraday dispersion function 3D cube. 
MRC 0600-399 has a complicated Faraday depth structure, and some of the Faraday depth features are in good agreement with the 2D structure on the celestial sphere. For the northern jet of MRC 0600-399, the Faraday depth values show a clear gap at the tip region. Based on the polarization analysis and comparison with the other observation, we concluded that the synchrotron radiation from the tip region of the northern jet originates from the magnetic field on the cold front.

\end{abstract}




\section{Introduction} \label{sec:intro}
The morphology of astrophysical jets launched from radio galaxies includes distinctive bending features in addition to the well-known bipolar type. Such bending jets are collectively referred to as Head-Tail (HT) radio galaxies, first mentioned by \citet{ryle1968}. HT radio galaxies are further classified into two main types based on the angle between the tails: Wide-Angle-Tail (WAT) galaxies, where the angle is between 90$^{\circ}$ and 180$^{\circ}$, and Narrow-Angle-Tail (NAT) galaxies, where it is between 0$^{\circ}$ and 90$^{\circ}$ \citep{owen1976}. Although there are exceptions, most of these objects are found within galaxy clusters \citep{mingo2019, silverstein2018, blanton2001, wing2013}. Notably, WATs are often found in merging galaxy clusters and are used as tracers for such events \citep[see reviews]{o'dea2023}. HT radio galaxies play a crucial role in understanding the dynamics of the intracluster medium (ICM) and galaxies within them \citep{hintzen1978, hintzen1984, blanton2000, smolcic2007, douglass2011, jones2017}.

Among HT radio galaxies, and particularly in WATs, the bending of jets has been primarily attributed to several mechanisms, including collisions with dense surrounding clouds \citep{higgins1999, wang2000, wiita2004}, buoyancy effects \citep{robertson1984, burns1981, burns1982, sakelliou1996}, electrodynamic effects \citep{eilek1984, fendt1998}, and ram pressure arising from ICM \citep{gomez1997a, gomez1997b}. Below, we briefly summarize the main features and limitations of each mechanism. See \citep{o'dea2023} for more details.
The hypothesis of jet bending due to collisions with dense ambient clouds, often referred to as jet-cloud interactions, has been suggested for several sources where such interactions are plausible. However, in many WATs, both jets exhibit bending at similar distances from the nucleus, and there is no compelling reason to expect dense gas clouds to be present at both of these locations simultaneously. Therefore, it is difficult to explain all WATs solely by this mechanism.

Buoyancy-induced jet bending occurs when the jet density is lower than that of the ICM. Although this mechanism can account for gently curved jets, it generally fails to explain cases where jets undergo abrupt changes in direction or sharp bends, as the buoyancy force typically produces only gradual curvature.

If the jets carry significant electric currents, they may experience Lorentz forces due to the presence of ordered magnetic fields within the galaxy cluster, potentially leading to jet bending. However, as with the dense cloud scenario, the requirement for ordered magnetic fields to exist at both bending locations is not necessarily justified, especially in cases where both jets bend at similar distances from the core. Furthermore, if the current flows along the jet direction and the same cluster-scale magnetic field is responsible for bending both jets, one would expect the bipolar jets to bend in opposite directions relative to the jet axis. Observationally, however, some WATs show both jets bending in the same direction, making it difficult to explain such sources with this mechanism.

When the ICM exhibits significant bulk motion, the jets may be bent by the resulting ram pressure. This effect is expected to be most pronounced near the cluster center, where the ICM velocity and density are highest, and indeed, many WATs located near cluster centers may be explained by this mechanism. However, in the outer regions of clusters, the ram pressure decreases, making this explanation less viable for WATs found at larger cluster-centric distances. In merging clusters, however, substantial bulk motions can persist even in the outskirts, so it remains possible that WATs could be formed by this mechanism in such environments. As discussed above, several mechanisms have been proposed to explain the formation of WATs. However, in recent years, objects have been discovered that exhibit bending trends that cannot be explained by any of these mechanisms.


Abell 3376 is a massive merging galaxy cluster with a redshift of $z$ = 0.046, having a mass of approximately 3.64$\times$10$^{14}$ $M_{\odot}$\citep{struble1999, girardi1998}. Its X-ray surface brightness and morphology suggest that it is undergoing a major merger \citep{machado2013, durret2013}. The radio galaxy MRC 0600-399, located within Abell 3376, is the second brightest cluster galaxy (BCG2). Unlike typical WAT galaxies, which exhibit C-shape or V-shape bends, MRC 0600-399's jets bend abruptly to the east after initially moving north-south \citep{bagchi2002, bagchi2006}. This bending is opposite to the direction of stripping that occurs with the ram pressure expected if a sub-cluster of BCG2 is going toward the east direction based on the global X-ray morphology.

Recent observations using MeerKAT have provided unprecedented detail of MRC 0600-399's structure, revealing a sharp 90-degree bend in its jets and the formation of ``double-scythe'' structures where some jet material propagates westward \citep{chibueze2021}. 
Three-dimensional magneto-hydrodynamic (MHD) simulations suggest that the 90-degree bending of the jet is caused by interactions with large-scale magnetic fields along cold fronts, assuming an average field strength of 10 $\mu$G and a jet kinetic luminosity of 2.5 $\times$ 10$^{43}$ erg s$^{-1}$\citep{chibueze2021}.
However, these simulations cannot fully explain the extremely long propagation distance ($\sim$ 100 kpc) of the northern jet after bending with the sharp shape, suggesting that additional physical mechanisms need to be considered.

This study aims to investigate the unexplored characteristics of MRC 0600-399 and its surroundings using MeerKAT data with polarization analysis. Although polarization studies of Abell 3376, including MRC 0600-399, have been conducted in the literature \citep{hu2024, gustafsson2024}, these focus on methodological developments rather than detailed discussions of MRC 0600-399's polarization properties or potential physical models. Therefore, this research will employ Faraday tomography, which is a modern technique of radio polarimetry (see \citealt{takahashi2023} for a review), to analyze the three-dimensional polarization structure of MRC 0600-399 and its environment. The paper is structured as follows: Section 2 describes the observations, data reduction and polarization analysis methods; Section 3 presents the analysis results; Section 4 discusses the findings; and Section 5 concludes the study.

\section{Data analysis}
\label{sec:obs}
\subsection{Observation and data reduction}
Abell 3376, which includes the radio galaxy MRC 0600-399, was observed on July 1, 2019, using 60 antennas of the MeerKAT array in the L-band (856 MHz to 1,712 MHz) (project ID: SCI-20190418-JC-01). For detailed telescope parameters, please refer to \citet{chibueze2021}.

For details on Stokes I calibration and imaging, refer to \citet{chibueze2021}. Here, we provide an overview:
J0408-6564 was used as the primary flux and bandpass calibrator, while J0616-3456 served as the secondary gain calibrator. General data reduction was performed using OXKAT\footnote{https://ascl.net/code/v/2627}, a semi-automatic pipeline. Imaging was carried out using WSClean \citep{offringa2014} with Briggs weighting and a robustness parameter of -0.3.
For wide-band imaging, we employed multi-frequency synthesis (MFS) in joined-channel deconvolution mode, with a central frequency of 1,283 MHz. The synthesized beam after imaging measured 5.8 arcsec $\times$ 5.5 arcsec, with a beam position angle (BPA) of 151.3 deg. To facilitate comparison with the polarization map discussed later, this MFS image was smoothed using CASA (ver 6.5.4.9), resulting in a final beam size of 7.2 arcsec $\times$ 7.1 arcsec and a BPA of 89.7 deg. The root-mean-square noise level was determined to be 7.8 $\mu$Jy per beam.

For details on the calibration of Stokes $Q$ and $U$, please refer to \citet{knowles2022} and OBIT DEVELOPMENT MEMO SERIES NO. 62\footnote{https://www.cv.nrao.edu/~bcotton/ObitDoc/MKPoln.pdf}.
The calibration of Stokes $Q$ and $U$ was initially performed using the SARAO calibration pipeline. This calibration utilizes the hardware noise diode calibration to measure the X-Y phases after the injection following the feeds.
Following the basic calibration, further calibration and imaging were conducted using the OBIT package \citep{cotton2008}. Due to MeerKAT's high stability, it is possible to reference calibration parameters obtained from a properly calibrated dataset using the same reference antenna \citep{plavin2020}. Based on this, the process determines the residual X-Y phases and the instrumental polarization terms. For MeerKAT, instrumental polarization signals become significant outside the half-power beam width of the primary beam. However, since our target is only about 0.09 degrees away from the phase center, we consider the impact of this effect to be very small.
Imaging of Stokes $Q$ and $U$ was performed using MFImage (wide-band, wide-field imager). To accommodate the high-frequency resolution required for the subsequent Faraday tomography polarization analysis, a maxFBW value of 0.01 was used. This resulted in the creation of 68 subband images. After removing planes with high noise levels, 47 subbands in the frequency range of 890--1672 MHz were ultimately used for analysis.
The final beam size is 7.2 arcsec $\times$ 7.1 arcsec, with a BPA of 89.7 deg. The root-mean-square noise levels for Stokes $Q$ and $U$ vary with frequency, ranging 9.8--24 $\mu$Jy beam$^{-1}$ and 27--48 $\mu$Jy beam$^{-1}$, respectively.

\subsection{Polarization analysis}
We performed Faraday tomography for polarization analysis \citep{burn1966, brentjens2005, akahori2018, ideguchi2018, takahashi2023}. Additionally, to achieve 3D visualization without the influence of unrelated foregrounds, we created a pseudo-3D cube \citep{rudnick2024}. Here, we describe the detailed procedure.

\subsubsection{RM-Synthesis and RM-Clean}
Faraday rotation measure synthesis (RM-Synthesis) is one of the most fundamental techniques in Faraday tomography \citep{brentjens2005}. In RM-Synthesis, the polarized intensity $P$(RA, Dec, $\lambda^2$), expressed as a function of the square of the observed wavelength, is Fourier transformed to obtain a three-dimensional cube of the Faraday dispersion function (FDF) $\tilde{F}$(RA, Dec, $\phi$). Here, $\phi$ [rad m$^{-2}$] is the Faraday depth, given by the following equation:
\begin{equation}
    \phi = 0.81 \int_{\rm observer}^{\rm source} n_e B_{||} dl
\end{equation}
Here, $n_e$ is a free electron density [cm$^{-3}$], $B_{||}$ is a magnetic field component parallel to the line of sight [$\mu$G], and $l$ is a path length along the line of sight [pc].

It should be noted that the FDF $\tilde{F}$ obtained from RM-Synthesis is the true FDF $F$ (RA, Dec, $\phi$) convolved with the RM spread function (RMSF). Therefore, RM-Clean, which operates on the same principle as CLEAN used in radio interferometry imaging, is employed to perform deconvolution of $\tilde{F}$ \citep{heald2009}. The FWHM of the RMSF can be expressed by the following equation using the longest observed wavelength $\lambda_{max}$ and the shortest wavelength $\lambda_{min}$:
\begin{equation}
    \delta\phi \approx \frac{2\sqrt{3}}{\lambda^2_{max} - \lambda^2_{min}}\sim 42.7\ {\rm rad\ m^{-2}}
\end{equation}
Deconvolution helps mitigate the effects of sidelobes in the Faraday spectra. In addition, the maximum observable Faraday depth can be expressed as:
\begin{equation}
    ||\phi_{\rm max}|| \approx \frac{\sqrt{3}}{\delta\lambda^2},
\end{equation}
where $\delta\lambda^2$ denotes the channel width in wavelength squared \citep{brentjens2005}. For the widest channel width of our dataset, this yields $||\phi_{\rm max}|| \sim 405$ rad m$^{-2}$.

We utilized RM-tools to perform RM-Synthesis and RM-Clean \citep{purcell2020}\footnote{https://github.com/CIRADA-Tools/RM-Tools}. RM-tools can automatically handle data with non-uniform sampling in the square of the wavelength. Our process consisted of the following steps. We first executed the "rmsynth3d" task using Stokes Q and U data to obtain the three-dimensional FDF $\tilde{F}$. Subsequently, we ran the "rmclean3d" task using the dirty map output from rmsynth3d. Here, we set the threshold value to 20 $\mu$Jy per beam. This value was chosen as five times the highest rms value in the $\phi$ plane of the off-source region in the dirty map, ensuring careful cleaning. We also set the gain value to 0.05.

\subsubsection{Pseudo-3D visualization}
\label{make_pseudo-3D}
The FDF obtained through RM-Synthesis or RM-Clean often includes information on structures unrelated to the target object, such as components from the Milky Way. While this FDF provides accurate information, it also introduces significant complexity in interpreting the polarization structure of the target. To address this, a powerful visualization technique has recently been developed, which is available at GITHUB\footnote{https://github.com/candersoncsiro/rmsynth3d}. For a detailed explanation, please refer to \citet{rudnick2024}. Here, we provide an overview of the method.
First, based on the three-dimensional cleaned FDF map obtained from RM-Clean, two-dimensional maps of peak intensity and corresponding Faraday depth at that peak are created. These can be generated using the ``rmtools\_peakfitcube" task in RM-tools. Next, a three-dimensional intensity map is reconstructed from these two types of maps. If we define the Faraday depth range of the output 3D cube as $\phi_1$ to $\phi_2$, the total number of pixels along the $\phi$ axis as $n_{\phi}$, and the pixel size along the $\phi$ axis as $\Delta\phi=\frac{\phi_2-\phi_1}{n_{\phi}}$, any coordinate in the reconstructed 3D cube can be expressed as follows:

\begin{equation}
    \mathbb{P}(i, j, k)=
\begin{cases}
    P(i,j)  & {\rm if\ \phi(i,j)=\phi_k\pm \frac{\delta\phi}{2}} \\
    0       & {\rm otherwise}
\end{cases}
,
\end{equation}

where (RA, Dec) is represented by the pixel coordinates $(i, j)$ and $\phi_k$ is the $\phi$ values at pixel $k$. This approach enhances the visibility of the local Faraday effects in the target object.

We set the Faraday depth range to -405 to 405 rad m$^{-2}$, aligning it with the cube obtained from RM-Synthesis. Additionally, we set $n_{\phi}$ = 90 and $\Delta\phi$ = 9 rad m$^{-2}$, which divides the FWHM of the RMSF ($\delta\phi \sim 42.7$ rad m$^{-2}$) into approximately five pixels. This choice of pixel size follows the standard practice in interferometric imaging, where the pixel size is typically set to approximately one-fifth of the beam size. Finally, we adopted a Faraday smoothing width of $\phi_{\mathrm{sm}} = 3$ rad m$^{-2}$, balancing between maximizing visibility and avoiding the creation of artifacts. We adopted a Faraday smoothing width of $\phi_{\mathrm{sm}} = 3$ rad m$^{-2}$. \citet{rudnick2024} notes that values in the range of $\phi_{\mathrm{sm}}=3-5$ typically maximize the visibility of the variations while avoiding artifacts due to oversampling. To validate this, we compared results for $\phi_{\mathrm{sm}} = 1,\ 3,\ $and 5. For $\phi_{\mathrm{sm}} = 1$, fine-scale structures became apparent, but diffuse emission was suppressed, and distinguishing genuine features from artifacts proved challenging. In contrast, $\phi_{\mathrm{sm}} = 3$ and $\phi_{\mathrm{sm}} = 5$ produced qualitatively similar results. However, the $\phi_{\mathrm{sm}} = 5$ case naturally produced broader features in the Faraday depth direction, complicating quantitative analysis of absolute Faraday spectrum widths. Therefore, we selected $\phi_{\mathrm{sm}} = 3$ rad m$^{-2}$ as a balance between detail preservation and artifact avoidance. All subsequent discussions in this paper focus on relative differences in Faraday spectrum widths rather than absolute values. Finally, we used SlicerAstro for 3D visualization \citep{punzo2015, SlicerAstro, punzo2017}.



\section{Results}
\subsection{Two-dimensional structures}
\label{result_2D}

\begin{figure*}
    \centering
    \includegraphics[width=1.0\linewidth]{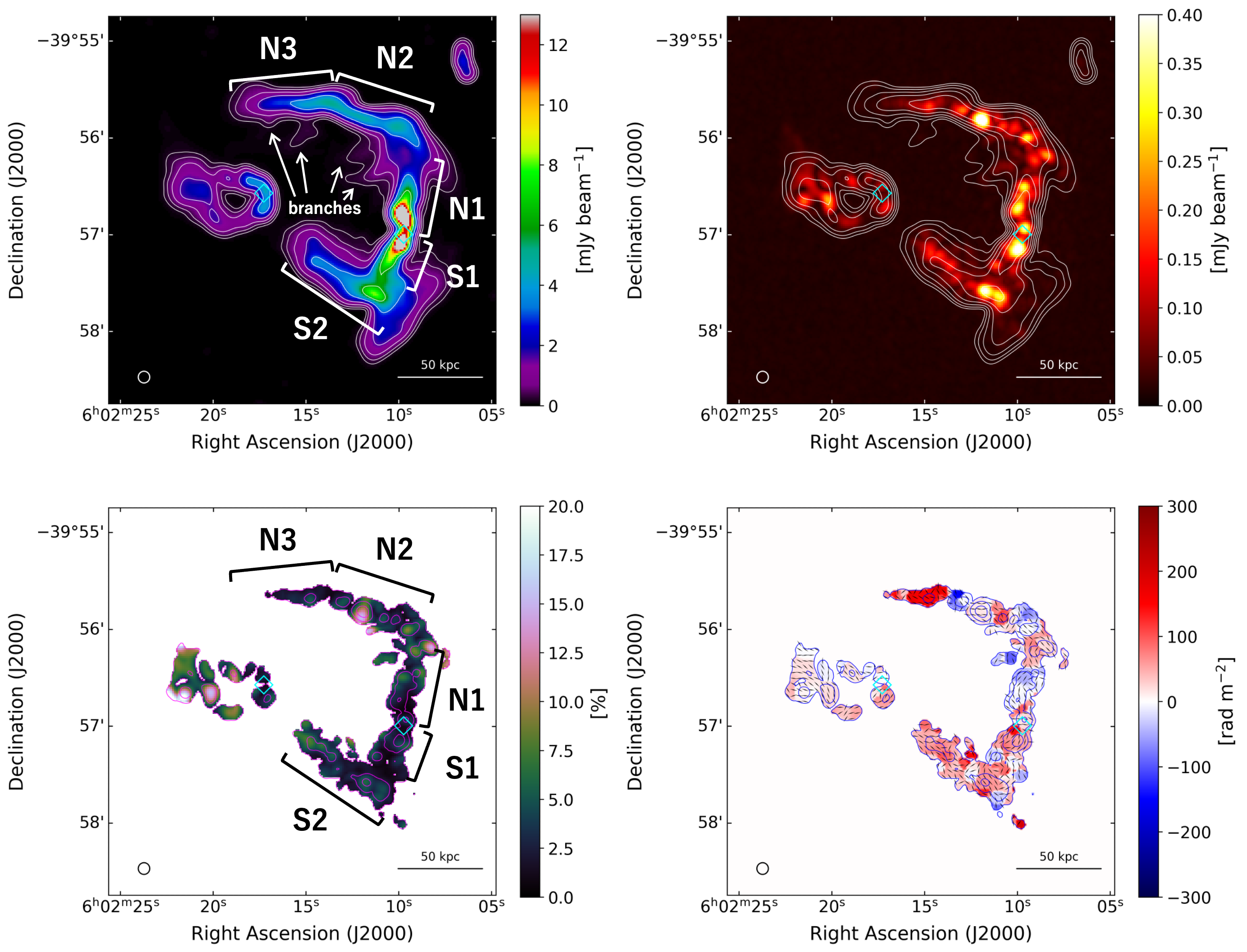}
    \caption{(Top-left) Wide-band total intensity map of MRC 0600-399 and Galaxy B at 1.28 GHz. The white contours show the intensities of rms $\times$ [25,50,100,200,400,800]. The cyan diamonds show the positions of the host galaxies. (Top-right) Wide-band polarized intensity map. The frequency range is 890--1678 MHz. The white contours are same with the total intensity map. (Bottom-left) Fractional polarization map. The magenta contours show the polarized intensities of rms $\times$ [3,9,24,45,90]. (Bottom-right) Peak Faraday depth map. The black ticks represent the intrinsic magnetic field vectors corrected for Faraday rotation effects. The blue contours are same with the fractional polarization map.}
    \label{fig:2Dmaps}
\end{figure*}


Here, we first show the two-dimensional structure projected onto the celestial sphere. The top-left panel of Figure \ref{fig:2Dmaps} shows a total intensity map centered at 1.28 GHz. We briefly describe some of the structures identified by \citet{chibueze2021}. Both the northern and southern jets of MRC 0600-399 bend sharply, with the main structure extending eastward while some gas flows to the west (double-scythe structures). After bending, the northern jet propagates for about 100 kpc (N2 and N3), while the southern jet extends for approximately 50 kpc (S2). Some branches can be seen for the northern jet after bending. Additionally, another head-tail radio galaxy exists to the east of MRC 0600-399, which we refer to as Galaxy B. The jet ejected from Galaxy B curves more gently compared to the jets of MRC 0600-399 and has a D-like shape. We also mention the shape of the jets before bending, which were not pointed out by \citet{chibueze2021}. The northern jet initially inclines toward the northeast immediately after ejection; however, it wiggles near the peak intensity region and propagates in a northwestern direction (N1). On the other hand, the southern jet is observed to propagate gently in a southeastern direction (S1).

The top-right panel of Figure \ref{fig:2Dmaps} displays a wideband (890--1678 MHz) polarized intensity map, with white contours representing total intensity. Our findings are as follows. Compared to the total intensity structure, the polarized intensity is notably patchy, with regions of low intensity tending to correspond to boundaries where the Faraday depth values change significantly (see the bottom-right panel of Figure \ref{fig:2Dmaps}). Immediately after jet ejection, there are intensity peaks in both the northern and southern directions. In the southern jet, the polarized intensity decreases after the initial peak but brightens again in the region where the jet bends. In this bending region, while the total intensity shows a single peak, the polarized intensity structure exhibits two peaks. In contrast, the northern jet shows several polarized intensity peaks before bending. A peak in the polarized intensity is also observed in the bending region, but it is offset from the total intensity peak.

The bottom-left panel of Figure \ref{fig:2Dmaps} shows a fractional polarization map. Note that fractional polarization values are only plotted for regions, at which total and polarized intensities exceed three times the rms values in off-source regions. MRC 0600-399 generally has low polarization, with an average value of 3.83\ \%. The highest fractional polarization is observed in the intermediate region after the northern jet bends, i.e. in N2 region, reaching 15.5\ \%. In comparison, Galaxy B has slightly higher polarization than MRC 0600-399, with an average of 6.98\ \% and a maximum value of 24.1\ \%. In Galaxy B, there is a tendency for the polarization fraction to be higher in the tail region compared to the jet ejection area, which is similar to the characteristic of WAT radio sources \citep{cotton2025}.


The bottom-right panel of Figure \ref{fig:2Dmaps} displays the peak $\phi$ map, showing the Faraday depth value corresponding to the highest Faraday spectrum amplitude for each pixel on the celestial sphere. For MRC 0600-399, the northern jet exhibits alternating positive and negative values of the peak $\phi$. After bending, the eastern propagating tip region of the northern jet exhibits rapid variations in values, ranging from a minimum of -157 rad m$^{-2}$ to a maximum of 202 rad m$^{-2}$. The southern jet, while having some negative value regions, predominantly shows positive values. Galaxy B tends to have smaller absolute $\phi$ values and less variation compared to MRC 0600-399. The contour shows the distribution of polarized intensity. Interestingly, the borders of the patchy structures correspond well to the transitions of the peak $\phi$ values. We will see this correspondence in the next subsection in more detail.

The bottom-right panel of Figure \ref{fig:2Dmaps} overlays magnetic field vectors, representing intrinsic magnetic field vectors corrected for Faraday rotation effects. Within a range of 10.5 kpc from the jet ejection region, the magnetic field is oriented east-west but subsequently tends to align parallel to the jet flow. The magnetic field vectors change direction as the jets bend. However, some magnetic field vectors are perpendicular to the jet flow immediately after bending in both the northern and southern jets. 
After bending, the intensity peak region shows well-aligned magnetic fields parallel to the jet flow. Galaxy B clearly exhibits magnetic field vectors well-aligned with its D-shape.


\subsection{Three-dimensional structures}
\begin{figure*}
    \centering
    \includegraphics[width=0.75\linewidth]
    {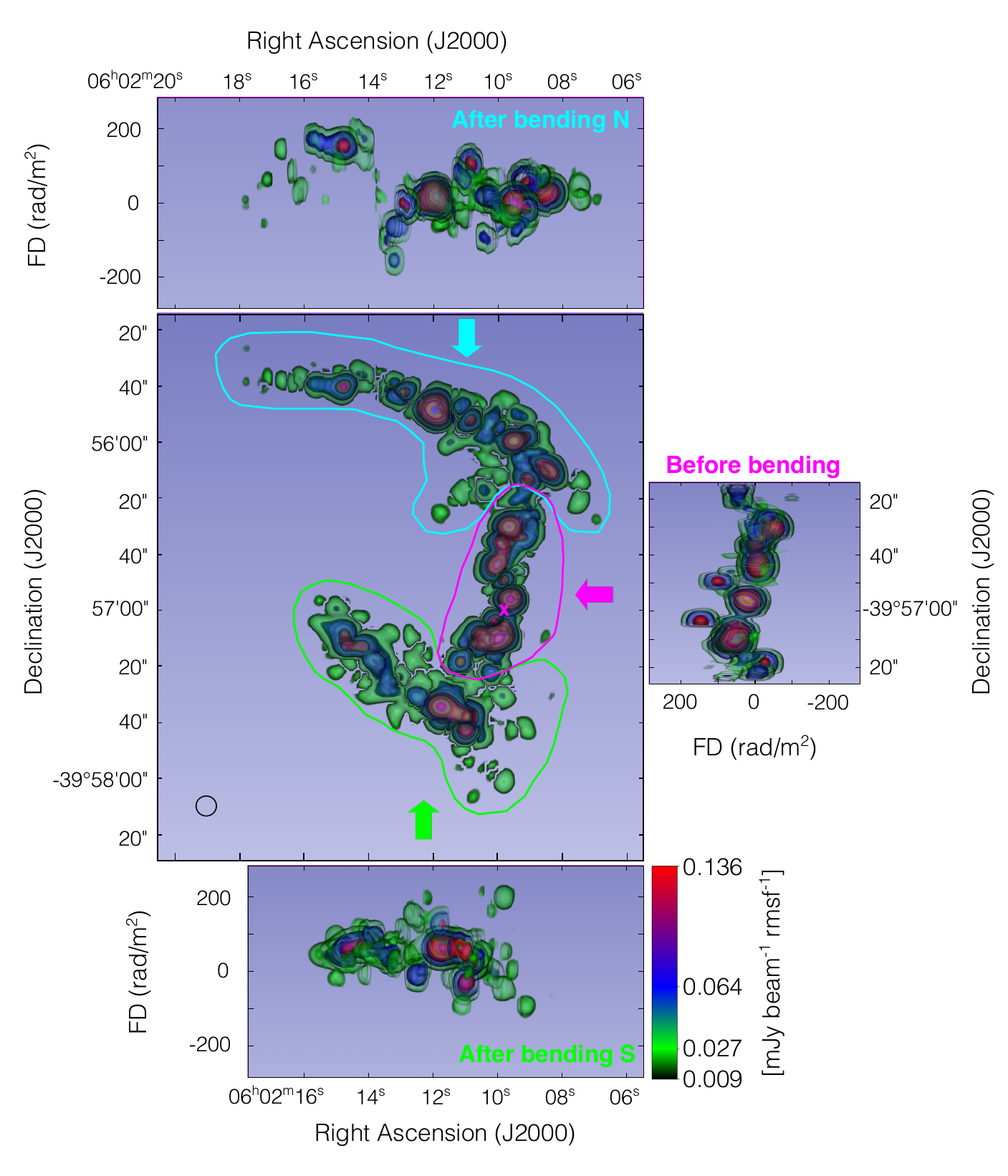}
    \caption{Pseudo-3D visualization map of MRC 0600-399. Middle-left panel shows the $\mathbb{P}$ distribution projected onto the celestial sphere. Top, middle-right, and bottom panels are side-views of the $\mathbb{P}$ from the directions of cyan, magenta, and light green arrows shown in middle-left panel, respectively. An animated version, which rotates in both the pitch and yaw directions, is available in the HTML version of the article. The animation lasts 20 seconds. The z-axis corresponds to the Faraday depth, with values increasing in the direction indicated by the arrow.}
    \label{fig:MRC3D}
\end{figure*}

\begin{figure*}
    \centering
    \includegraphics[width=0.65\linewidth]{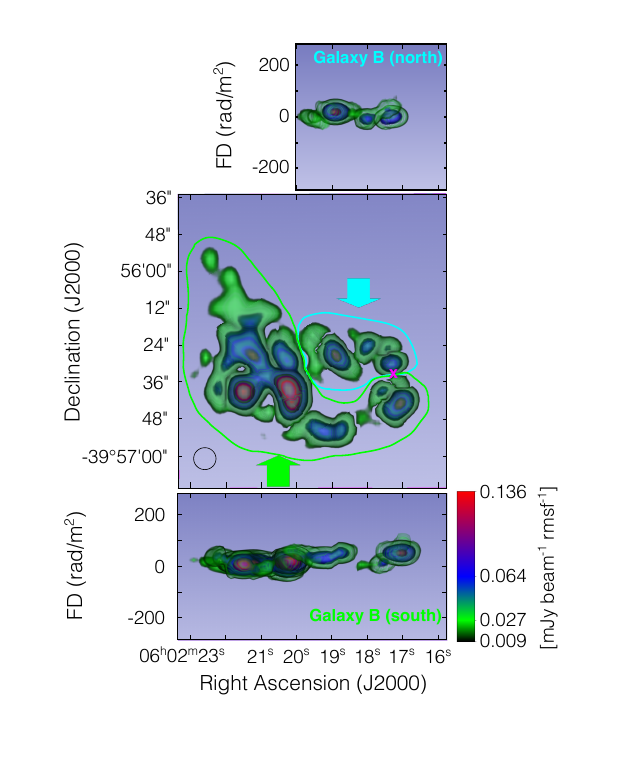}
    \caption{Similar to Figure \ref{fig:MRC3D} but for Galaxy B. An animated version, which rotates in both the pitch and yaw directions, is available in the HTML version of the article. The animation lasts 20 seconds. The z-axis corresponds to the Faraday depth, with values increasing in the direction indicated by the arrow.}
    \label{fig:galB3D}
\end{figure*}
Here, we explain the structures observed in the Pseudo-3D visualization map created in Section \ref{make_pseudo-3D}. Note that for clarity, the Pseudo-3D visualization maps of MRC 0600-399 and Galaxy B have been separated.

Figure \ref{fig:MRC3D} shows the Pseudo-3D visualization map of MRC 0600-399. The central panel represents the projection onto the celestial sphere. Similar to the polarized intensity in the top-right panel of Figure \ref{fig:2Dmaps}, it exhibits a patchy structure. The top, bottom, and right panels of Figure \ref{fig:MRC3D} are maps viewed from the direction of the arrows, showing only the regions enclosed by cyan, light green, and magenta lines in the central figure, respectively. The vertical axis in the top and bottom figures and the horizontal axis in the right figure represent Faraday depth ($\phi$). An animation rotating the Pseudo-3D visualization map is available in the Supplemental data.

First, we focus on the structure of the jet before bending, enclosed by a magenta line in the central panel of Figure \ref{fig:MRC3D}. This region covers N1 and S1 regions shown in Figure \ref{fig:2Dmaps}. As mentioned in Section \ref{result_2D}, the northern jet initially propagates toward the northeast after ejection, then changes the direction to the northwest. In other words, it is the jet wiggles discussed in \citet{nakamura2001}, \citet{kigure2004}, and \citet{uchida2004}. The southern jet, on the other hand, is observed to move gently to the southeast after ejection. Similar trends can be confirmed in the Pseudo-3D visualization map (central panel of Figure \ref{fig:MRC3D}). Interestingly, the wiggle of the jet can also be observed in the Faraday depth space, as shown in the right panel of Figure \ref{fig:MRC3D}. We will discuss the wiggle structure in Section \ref{discussion:before-bend} in more detail.
%

Next, we focus on the northern jet after bending, enclosed by a cyan line in the central panel of Figure \ref{fig:MRC3D} (the top panel of Figure \ref{fig:MRC3D}). This region covers N2 and N3 regions shown in Figure \ref{fig:2Dmaps}. Compared to the jet before bending, we can see the presence of faint and diffuse emission. Additionally, many regions have multiple Faraday components along each right ascension slice, which have different $\phi$ peak values. This implies the existence of multiple polarized radiation sources in the area. These different polarized components could be aligned along the line of sight, or they could be located at different positions on the plane of the sky. We will discuss these structure in Section \ref{discussion:after-bend}. Immediately after bending, the jet exhibits a complex structure in Faraday depth space, although it converges to a single peak near right ascension 06$^{\rm h}$02$^{\rm m}$12$^{\rm s}$. Subsequently, we can observe the peak Faraday depth rapidly transitioning to around -150 rad m$^{-2}$, followed by another abrupt change close to 150 rad m$^{-2}$. We can see the clear discontinuity in the region.

Finally, we focus on the southern jet after bending, enclosed by a light green line in the central panel of Figure 3 (the bottom panel of Figure 3). Similar to the northern jet after bending, the southern jet also exhibits multiple Faraday components at each right ascension slice. However, unlike the northern jet, there is no observable shift in peak Faraday depth extending to several hundred rad m$^{-2}$. It means that the southern jet after bending is a similar structure with the northern jet before converging to a single $\phi$ peak.

Figure \ref{fig:galB3D} shows the Pseudo-3D visualization map of Galaxy B. Similar to Figure \ref{fig:MRC3D}, the central panel shows a celestial sphere projection, while the top and bottom panels display the cyan and light green outlined regions viewed from the directions indicated by the arrows. The vertical axes of the top and bottom maps represent Faraday depth. Galaxy B's jets exhibit significantly smoother Faraday spectral structure compared to MRC 0600-399, with a smaller variation in Faraday depth. In addition, the jets have a single component along the each right ascension slice. It emphasizes the complexity of MRC 0600-399 compared to Galaxy B.

\begin{figure*}
    \centering    \includegraphics[width=0.9\linewidth]{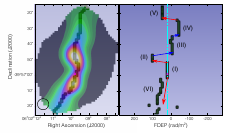}
    \caption{(Left) Total intensity map of the jets before bending. The overlaid boxes represent the pixel positions of the intensity peak at each declination. (Right) Faraday depth distribution of each intensity peak position. The cyan dashed line shows the foreground Galactic rotation measure in the target direction.}
    \label{fig:stream}
\end{figure*}

\section{Discussion}
\subsection{Wiggle of jet before bending}
\label{discussion:before-bend}

\begin{figure}
\includegraphics[width=1.0\linewidth]{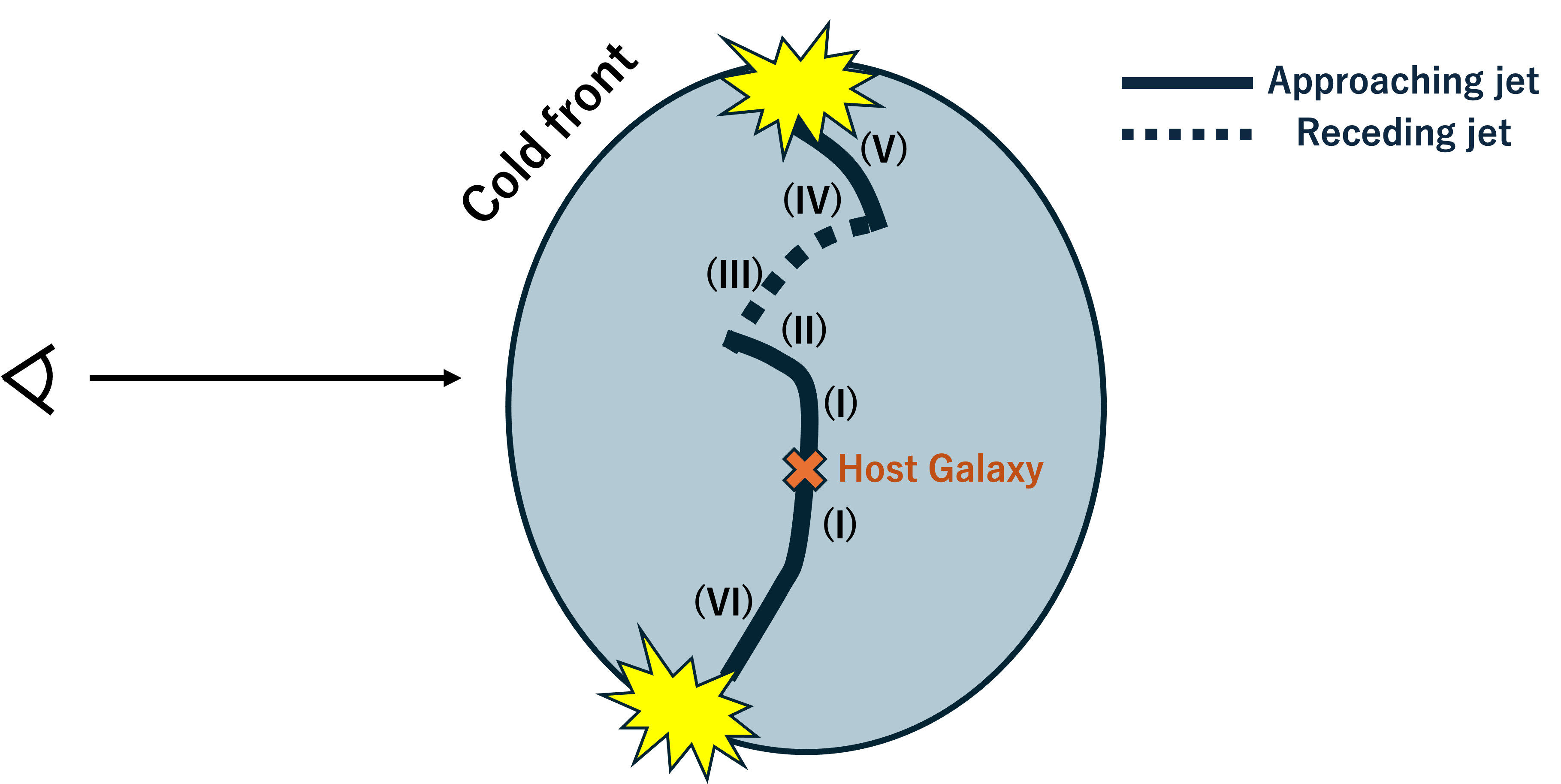}
    \caption{Schematic drawing of the MRC 0600-399 jets before bending. An observer is looking at the jets from the left side of the figure. Roman numerals in the panel correspond to the one in Figure \ref{fig:stream}.}
    \label{fig:stream_schematic}
\end{figure}
This section examines the relationship between the jet structures before bending projected onto the celestial sphere (N1 and S1 in Figure \ref{fig:2Dmaps}) and their corresponding Faraday depth distributions. As mentioned in Section \ref{result_2D}, the northern jet initially propagates northeast before altering its trajectory at the peak of the total intensity to move northwest. In contrast, the southern jet propagates smoothly southeast without abrupt directional changes.

Such wiggled structures are observed not only in head-tail radio galaxies but also in typical AGN jets {\citep{horton2025}. Possible causes include precession of the jet ejection region, Rayleigh-Taylor instability, and current-driven kink instability. The magnetic field configuration associated with these jet wiggles has been discussed in \citet{nakamura2001}, \citet{kigure2004}, and \citet{uchida2004}. These studies propose that initially helical magnetic fields surrounding the jet become aligned with the wiggling flow pattern while maintaining their twisted configuration. Furthermore, theoretical models predict that such structures should exhibit characteristic gradients in Faraday rotation measures along the wiggled jet path. 

Figure \ref{fig:stream} shows how Faraday depth values evolve with the jet’s trajectory. The left panel displays the total intensity distribution of the jet before bending, with overlaid boxes indicating pixel positions where the total intensity reaches its maximum in each declination. The right panel plots Faraday depth values at these box positions, with the x-axis representing Faraday depth. The vertical dashed cyan line marks the average foreground Faraday depth from the Milky Way ($\phi_{MW} \sim 21.76$ rad m$^{-2}$) \citep{hutschenreuter2022}.

Under the assumption that the intervening Faraday screen consists solely of the Milky Way and the intracluster medium (ICM) of Abell 3376, and that the turbulent ICM contributes only to the dispersion of the Faraday spectrum (not the peak $\phi$), regions with Faraday depths exceeding $\phi_{MW}$ indicate dominant line-of-sight magnetic fields toward the observer, while regions with lower values suggest fields receding from the observer.

For the northern jet (N1), Faraday depth variations correlate with structural changes in its celestial projection. Initially, near the jet ejection region, Faraday depths are comparable to $\phi_{MW}$ ($\sim$ 18 rad m$^{-2}$, I in Figure 5), implying negligible line-of-sight magnetic fields. At the total intensity peak, Faraday depth rises sharply to $\sim$ 99 rad m$^{-2}$ (II). As the jet turns northwest, values drop below $\phi_{MW}$ ($\sim$ -9 rad m$^{-2}$, III), reaching a minimum of $\sim$ -45 rad m$^{-2}$ further along the trajectory (IV). Before bending, Faraday depth increases again to $\sim$ 45 rad m$^{-2}$ (V). For the southern jet, Faraday depths gradually increase from 45 to 81 rad m$^{-2}$ (VI), consistent with its smooth southeastward propagation and lack of abrupt directional changes. Minor outliers may reflect jet bending effects.


Based on the projected structure on the celestial sphere and the Faraday depth distribution, we may restore the jets' structure in the direction of the line of sight. Note that this analysis relies on two key assumptions mentioned above: first, that the intervening Faraday screen between the jet and the observer consists solely of the Milky Way Galaxy and the intracluster medium (ICM) of Abell 3376; second, that the ICM is turbulent, with its contribution to Faraday rotation manifesting only in the dispersion of the Faraday spectrum, not the peak $\phi$ values. As shown in Figure \ref{fig:2Dmaps}, prior to jet bending, the magnetic field vectors align with the jet flow direction, suggesting that the large-scale magnetic field associated with the jet is parallel to its flow. When the jet’s flow direction changes along the line of sight, the direction of the magnetic field should also be changed.

Figure \ref{fig:stream_schematic} is the schematic drawing of the jet's trail in the direction of the line of sight. For the northern jet, the structure evolves as follows: From the jet ejection region to the total intensity peak (I to II), the jet transitions from a direction perpendicular to the line of sight to one approaching the observer. 
An increase in total intensity, which is likely attributable to Doppler boosting, is observed in the region where the Faraday depth is greatest and the jet is most inclined toward the observer (II).
On the celestial sphere, the jet changes the propagating direction from northeast to northwest at the total intensity peak, with the Faraday depth changing from positive to negative, indicating a reversal in the line-of-sight direction from approaching to receding (II to III). A smooth transition in the jet’s line-of-sight direction would theoretically produce a region with Faraday depth values comparable to $\phi_{MW}$ during the sign reversal, but this feature is not observed, likely due to insufficient spatial resolution. Subsequently, the jet propagates northwestward while receding along the line of sight (III to IV). 
In this region, a decrease in total intensity is observed, which can be attributed to the jet receding from the observer.
Upon reaching the bending region, the jet’s trajectory likely redirects back toward the observer, where the jet change the propagating direction again on the celestial sphere (IV to V). 
Taking into account the morphology on the celestial sphere, it is highly probable that the northern jet is undergoing precession.
For the southern jet, the structure differs: On the celestial sphere, the jet gradually extends southeastward from the jet ejection region, with the Faraday depth increasing in this region, suggesting a gradual approach along the line of sight (I to VI). 

These observations enable a discussion of the three-dimensional jet structure through Faraday depth distributions. However, the validity of this interpretation depends critically on the stated assumptions.

\subsection{Polarization characteristics of the northern jet after bending}
\label{discussion:after-bend}
\begin{figure}
    \centering
    \includegraphics[width=1.0\linewidth]{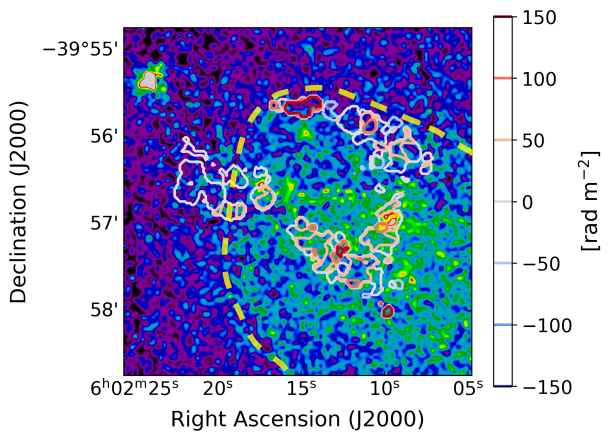}
    \caption{XMM-Newton image around MRC 0600-399 and Galaxy B. See \citet{chibueze2021} for the observation details. The contours represent the peak Faraday depth distribution shown in the bottom-right panel of Figure \ref{fig:2Dmaps}, and the color of each contour corresponds to the value indicated on the color bar. A yellow dashed line indicates the position of the cold front.}
    \label{fig:x-ray}
\end{figure}


We discuss the polarization characteristics of the bent structure in the northern jet of MRC 0600-399 beyond its bending point (N2 and N3). As shown in the top panel of Figure \ref{fig:MRC3D}, immediately after the jet bending, multiple Faraday components appear to coexist, exhibiting a significantly broader structure in Faraday depth compared to Galaxy B. Based on the model proposed by \citet{chibueze2021}, where jet bending occurs due to collision with the magnetic field along the cold front, this phenomenon can be explained by the coexistence of the main jet flow and backflow gas compressed at the bending interface. 

The northern jet shows a peak in polarized intensity near right ascension 06$^{\rm h}$02$^{\rm m}$12$^{\rm s}$ and peak Faraday depth of 0 rad m$^{-2}$ (Figures \ref{fig:2Dmaps} and \ref{fig:MRC3D}). This region is located within N2, where spectral flattening occurs and implies the cosmic ray electron re-acceleration \citep{chibueze2021}. The increase in polarized intensity in this region supports this hypothesis. From the end of N2 to N3, the peak Faraday depth value shifts rapidly to -150 rad m$^{-2}$ and then close to +150 rad m$^{-2}$. The absolute values of peak Faraday depth in this region are larger compared to other areas. High Faraday depth values typically indicate high thermal electron density and/or strong line-of-sight parallel magnetic field. The X-ray intensity distribution around MRC 0600-399 shows lower thermal electron density in this region compared to the area before the jet bends (Figure \ref{fig:x-ray}). This suggests the presence of a strong line-of-sight parallel magnetic field in this region.

Here, we investigate the cause of the prominent line-of-sight parallel magnetic field and its reversal from the end of N2 to N3. The first plausible scenario involves an abrupt bending of the jet along the line of sight. As assumed in Section \ref{discussion:before-bend}, the jet contains magnetic fields aligned with its flow. If the jet bends sharply away from the observer near the N2-N3 boundary and subsequently bends back toward the observer within N3, this configuration could explain the abrupt shift in Faraday depth values. However, this interpretation faces challenges. For instance, if wiggling analogous to the ``jet before bending'' occurs, an intermediate region with Faraday depth values near 0 rad m$^{-2}$ would be expected during the polarity reversal. In contrast, the observed depolarization in the intermediate region between N2 and N3 instead suggests an abrupt magnetic field reversal rather than a gradual transition. This depolarization implies that the magnetic field orientation changes too rapidly for coherent polarization signals to persist across the transition zone.

The second possibility is that the polarized emission observed in N3 originates not from synchrotron radiation associated with jet magnetic fields but rather from the one associated with large-scale magnetic fields in the cold front.
Note that we assume the supply source of the cosmic-ray electrons (CREs) is the northern jet after bending. The magnetic fields associated with the cold front exist on its surface. As discussed in \citet{chibueze2021}, we propose that magnetic tension from these fields could push back and bend the jet structure. Crucially, the magnetic fields need not strictly align with the east-west direction of the bent jet propagation. Since the jet deflection only requires magnetic field presence at the collision interface, the cold-front magnetic fields may possess components along the line of sight in addition to the celestial east-west orientation. 
If the N3 structure originates from the magnetic fields on the cold front, the observed sharp Faraday depth variation could indicate such line-of-sight magnetic components. 

The cold-front magnetic fields creating N3's main structure could exist as either a continuous structure or multiple discrete strands. In the latter case, energy loss through cosmic-ray electron diffusion between magnetic strands must be considered. This scenario aligns with observed rapid intensity decrease toward its tip, making it a plausible explanation.

\subsection{Cosmic-ray electrons propagation}
To support this interpretation, however, two key questions must be addressed: (1) Can CREs jump from the jet-associated magnetic fields to the cold front-associated magnetic fields? (2) After the transfer occurs, can these electrons propagate over a distance of approximately 50 kpc? 

Regarding question (1), CRE transfer from jet magnetic fields to cold front magnetic fields can be explained by two mechanisms: (a) perpendicular diffusion of cosmic-ray electrons and (b) magnetic reconnection. For (a), assuming a jet magnetic field strength of 10 $\mu$G \citep{chibueze2021}, electrons emitting synchrotron radiation at 1.28 GHz would have an energy of approximately 2.8 GeV. The Larmor radius for such electrons would be around 0.3 $\mu$pc. If magnetic field perturbations exist on spatial scales smaller than this value, perpendicular diffusion could enable CREs to transition between these two magnetic field systems. 
For (b), magnetic reconnection could occur at locations where jet and cold front magnetic fields interact, such as after bending in the northern jet. In this case, CREs might be reaccelerated within reconnection regions, which could also explain spectral index flattening observed in N2. 

For question (2), mechanisms enabling CRE propagation over $\sim$50 kpc after transferring to cold front magnetic fields include (a) parallel diffusion of cosmic-ray electrons, (b) cosmic-ray streaming (CR streaming), and (c) thermal conduction. For (a), considering electron energy and magnetic field strength, an unusually high diffusion coefficient on the order of $\sim$10$^{32}$ cm$^{2}$/s would be required—a few orders of magnitude larger than values estimated from cosmic-ray observations in our Galaxy or theoretical models for galaxy clusters \citep{strong2007, fang2016}. However, downstream regions of cold fronts may exhibit reduced turbulence compared to cluster centers, potentially allowing for higher diffusion coefficients aligned with large-scale magnetic fields. 
For case (b), we consider the cold-front magnetic fields with components along the line of sight from the end of N2 to N3. The maximum magnetic coherence scale is estimated to be approximately 25 kpc. On this scale, the Faraday depth varies by about 300 rad m$^{-2}$. Assuming a magnetic field strength of 10 $\mu$G, the corresponding electron number density is estimated to be 1.5 $\times$ 10$^{-2}$ cm$^{-3}$. Using these parameters, the Alfv\'{e}n velocity is calculated to be approximately 1800 km/s. Consequently, the propagation time for cosmic-ray electrons (CREs) along N3 is about 265 Myr. In contrast, the cooling timescale can be estimated following \citet{nishiwaki2022}. Since the spectral break frequency is not observed within our target frequency range, we adopt the lowest observed frequency, 890 MHz, as the break frequency, yielding a lower limit for the cooling time of 41 Myr. If the break frequency is as low as 1 MHz, the cooling time approaches 265 Myr. However, \citet{chibueze2023} suggest, based on lower-resolution observations, that the break frequency is likely higher than 88 MHz. Therefore, scenario (b) is considered unrealistic.
Finally, for (c), thermal conduction may play a role since it is more efficient along magnetic field lines than perpendicular to them. After interacting with cold front magnetic fields, jet gas might propagate along these fields due to thermal conduction effects.

\subsection{Characteristics of Galaxy B}
\label{discussion:galaxyB}
Finally, we consider the polarization characteristics of Galaxy B. The corresponding object is likely 6dFGS gJ060217.3-395635, with a redshift of z$\sim$ 0.04813, placing it behind the center of Abell 3376 \citep{moretti2017}. Unlike MRC 0600-399, the jet bending of Galaxy B is gradual, exhibiting a D-like shape. As mentioned in Section \ref{result_2D}, considering the polarization fraction trends, Galaxy B appears to be a typical WAT object. Its jet bends eastward, suggesting that Galaxy B was originally a member galaxy of the eastern cluster before the collision of two galaxy clusters and is experiencing ram pressure from west to east.

The lower panel of Figure \ref{fig:galB3D} shows a broadening of the widths of the Faraday component in east of the right ascension 06$^{\rm h}$02$^{\rm m}$19$^{\rm s}$.5, where the northern and southern jets overlap. Typically, when different polarized sources overlap along the line of sight, Faraday rotation from the foreground structure is expected, resulting in multiple Faraday components in the Faraday spectrum. If the resolution of the Faraday spectrum is improved, it may be possible to separate these multiple Faraday components. However, at least based on the present results, we find that the peak Faraday depth values of the Faraday components associated with the northern jet and the southern jet are not separated by more than $\delta \phi$. To explain this situation, we assume that the jet possesses a well-aligned magnetic field along its flow, that we are viewing the jet almost side-on, and that the jet bending occurs primarily in a plane perpendicular to the line of sight. Under these conditions, the foreground jet may not cause significant Faraday rotation. Consequently, the two Faraday components are likely to have similar peak Faraday depth values and appear to be degenerate. This unique configuration explains the observed broadening of Faraday component line widths without distinct separation of components, contributing to our understanding of Galaxy B's structure and its interaction with the surrounding cluster environment.

\section{Conclusion}
In this paper, we performed a polarization analysis of MRC 0600-399, the BCG2 of the merging galaxy cluster Abell 3376, as well as a nearby WAT radio source (Galaxy B). For the polarization analysis, we employed RM synthesis and further conducted three-dimensional visualization of the results (Pseudo-3D visualization). MRC 0600-399 exhibited a highly patchy distribution of polarized intensity, and, except for certain regions in the northern jet after bending, the source generally showed a low polarization fraction. In contrast, Galaxy B displayed a tendency for higher polarization fractions compared to MRC 0600-399. The distribution of Faraday depth values in MRC 0600-399 was found to correspond closely to the patchy structure seen in the polarized intensity. Overall, reversals in the sign of the Faraday depth were observed, and particularly large jumps in Faraday depth were evident in the northern jet after bending (N2 and N3). For the jets before bending, the sign-reversal structure in Faraday depth corresponded well with the wiggling structure seen in total intensity on the celestial sphere, suggesting that the Faraday depth reflects the line-of-sight structure of the jets before bending. The abrupt changes in Faraday depth values in the northern jet after bending (N2 and N3) can be explained by considering that the synchrotron emission in these regions originates from magnetic fields on the cold front of the galaxy cluster. There are several possible physical mechanisms by which cosmic-ray electrons, accelerated by the jet, could escape the jet’s magnetic field, transfer to the magnetic field on the cold front, and subsequently propagate over a distance of approximately 50 kpc. The Faraday depth structure of Galaxy B was found to be distributed within a narrower range and to have fewer Faraday components compared to MRC 0600-399.


\begin{acknowledgements}
We are grateful to Drs. C. S. Anderson, A. Seta, K. Kurahara and Ms. S. Bradbury for helpful discussion.
We thank the anonymous referee for useful comments and constructive suggestions.
This work was supported by JSPS KAKENHI Grant Numbers HS: 22K20386, MM: 22H01272, 23K22543, 24K00672, TTT: 21H01128, 24H00247.
TTT has also been supported in part by the Collaboration Funding of the Institute of Statistical Mathematics ``Machine-Learning-Based Cosmogony: From Structure Formation to Galaxy Evolution''. 
JOC acknowledges support from the Italian Ministry of Foreign Affairs and International Cooperation (MAECI Grant Number ZA18GR02) and the South African Department of Science
and Technology’s National Research Foundation (DST-NRF Grant
Number 113121) as part of the ISARP RADIOSKY2020 Joint
Research Scheme.

The MeerKAT telescope is operated by the South African Radio Astronomy Observatory, which is a facility of the National Research Foundation, an agency of the Department of Science, Technology and Innovation.

\end{acknowledgements}


\bibliography{sample631}{}
\bibliographystyle{aasjournal}



\end{document}